\documentclass[%
 preprint,
 amsmath,amssymb,
 aps,
prl,
floatfix,
]{revtex4-2}

\usepackage{tabularx}
\usepackage{graphicx}
\usepackage{dcolumn}
\usepackage{bm}
\usepackage{hyperref}
\usepackage{siunitx}
\usepackage{svg}
\usepackage{subfigure}

\usepackage{epstopdf}
\AppendGraphicsExtensions{.tif}

\usepackage[mathlines]{lineno}

\begin{document}


\title{Enhanced Hot Electron Preheat Observed in Magnetized Laser Direct-Drive Implosions}

\author{M. Cufari}
\email{mcufari@mit.edu}
\author{M. Gatu Johnson}
\author{C. K. Li}
\author{J. A. Frenje}%
 \affiliation{%
Massachusetts Institute of Technology Plasma Science and Fusion Center, Cambridge MA, 02139
}%

\author{P. W. Moloney}
\author{A. J. Crilly}
\affiliation{Blackett Laboratory, Imperial College, London SW7 2AZ, United Kingdom}

\author{P. V. Heuer}
\author{J. R. Davies}
\affiliation{Laboratory for Laser Energetics, University of Rochester, Rochester NY, 14623}

\date{\today}

\begin{abstract}
Hard x-ray emission, associated with hot electron preheat, in direct-drive implosions was observed to be enhanced by a factor of $1.5\pm0.1$ by application of a $10$ T magnetic field. The applied magnetic field reaches a quasi steady-state aligned with the ablation flow prior to the onset of laser-plasma instabilities in the corona. Hot electrons that would otherwise escape the corona and lead to capsule charging in unmagnetized implosions are confined in a mirror-mode of the magnetic field in magnetized implosions. These hot electrons are shown to subsequently pitch-angle scatter from the mirror onto the capsule, thereby leading to the observed hard x-ray generation in magnetized implosions. Consequently, the energy of charged-fusion products, associated with the capsule charging, are observed to decrease when the implosion is magnetized. These results intensify the need to mitigate laser-plasma instabilities -- particularly for magnetized implosions -- to maximize fusion gain and implosion efficiency. 
\end{abstract}
                         
\maketitle

\textit{Introduction} Laser driven inertial confinement fusion (ICF) uses lasers with intensities $I \sim 10^{14} - 10^{15}$ \si{\watt/\cm^2} to compress spherical capsules of $\sim 1$ milligram of deuterium-tritium fuel to densities $\gtrsim 100$ \si{\gram/\cm^3} and temperatures $\sim 10$ \si{\kilo\electronvolt} over timescales of $\sim~1 - 10$ \si{\nano\second} \cite{nuckolls_1972}. This is accomplished by laser-driven ablation of a spherical-shell, often composed of glass \cite{rosen_larsen_nuckolls_1977}, plastic \cite{nikroo1999}, or high-density carbon \cite{Biener_2009,ross2015}. Two approaches have been used historically to accomplish this: indirect-drive \cite{lindl_1995}, where lasers are incident on the interior of a radiation-cavity termed a `hohlraum', and direct-drive \cite{craxton_2015}, where lasers are incident on the capsule surface. This letter concerns a variation of the direct-drive approach where the capsule is immersed in an external magnetic field. The goal of magnetization of the implosion is to reduce electron thermal-conduction losses during the implosion, thereby achieving higher temperatures and greater thermonuclear fusion performance and/or reducing the required laser energy to achieve large energy gain. 

A major concern for the viability of direct-drive fusion at ignition energy scales, $E_{\rm{laser}} \gtrsim 1$ MJ, is the generation of suprathermal electrons, or hot electrons, via laser-plasma instability (LPI), that prematurely heat the fuel. Electron preheat reduces the compressibility of the fuel and can therefore be detrimental to ICF implosions as small changes in the density lead to large changes in the fusion yield, $E_{\rm{fusion}} \propto \rho^2$ \cite{smalyuk_2008, craxton_2015, cao_2022}. Magnetization of direct-drive implosions has hitherto been conjectured to reduce the amount of preheat via mitigation of LPI that generates hot electrons \cite{winjum_2018,xxli_2025}. However, we observe that the hot electron preheat is systematically increased by a factor $1.5\pm 0.1$ in magnetized direct-drive fusion experiments. This result presents a concern for the long-term viability of magnetized direct-drive implosions for high-yield/high-gain applications, unless laser plasma instabilities in the corona can be avoided, e.g., via the use of broadband lasers \cite{froula_2025}. The remainder of this letter is dedicated to the presentation of these data and the explanation of the origin of the enhanced preheat. 

\textit{Experiments} Magnetized direct-drive implosions were carried out at the OMEGA laser using glass capsules with $2.6\pm0.2$ \si{\micro\meter} thick shells and outer diameters of $850\pm 10$ \si{\micro\meter}. The capsules were filled with a mixture of deuterium ($\rm{P}_{\textrm{D}_2}\sim 14.7$ atm) and helium-3 ($\rm{P}_{\rm{He}^3} \sim 3$ atm). The implosions were driven using 60 laser beams pointed towards target chamber center to produce a spherically uniform irradiation pattern. The laser intensities were $\sim 1.0-1.2 \times 10^{15}$ \si{\watt/\cm^{2}} achieved utilizing $27 $ \si{\kilo\joule} delivered to the capsule in a $1$ \si{\nano\second} square-shaped pulse. Implosions were performed in which a pair of magneto-inertial fusion electrical discharge system (MIFEDS)\cite{fiskel_2015} coils were utilized to provide a $\sim 10$ \si{\tesla} axial magnetic field; additional implosions were performed without the magnetic field for reference -- note that in the unmagnetized implosions, the coils used to provide the magnetic field were not inserted into the target chamber and potential line-of-sight obstructions due to the coils in magnetized experiments are considered and ruled out.

In each experiment, the flux of hard x-rays $h\nu\gtrsim 60$ \si{\kilo\electronvolt}, and $h\nu\gtrsim 80$ \si{\kilo\electronvolt}, associated with a hot electron population producing bremsstrahlung radiation upon interaction with the ion-species in the capsule shell, were measured using the hard x-ray detector (HXRD) \cite{stoeckl_2001}. The energy spectra of charged particles produced by D$^3$He and DD reactions were measured using a standard suite of charged-particle spectroscopy diagnostics on OMEGA including wedge range filters (WRFs) and the charged-particle spectrometers (CPS1 and CPS2) \cite{seguin_2003}. These diagnostics provide detailed information on the time and volume integrated plasma conditions in the capsule core as well as information about electric fields/potentials that may be established during the laser drive due to the production/ejection of hot electrons from the coronal plasma. Such fields lead to systematic up-shifts in the energy spectrum of charged fusion products as they gain energy equivalent to the potential difference between the capsule and the detector\cite{hicks_2000}. Often, measured charged particle median energies in similar implosions are observed to lie $100$'s of keV above the birth energy of the particle due to the capsule potential.

\begin{figure}[p!]
    \includegraphics[width=0.9\linewidth]{"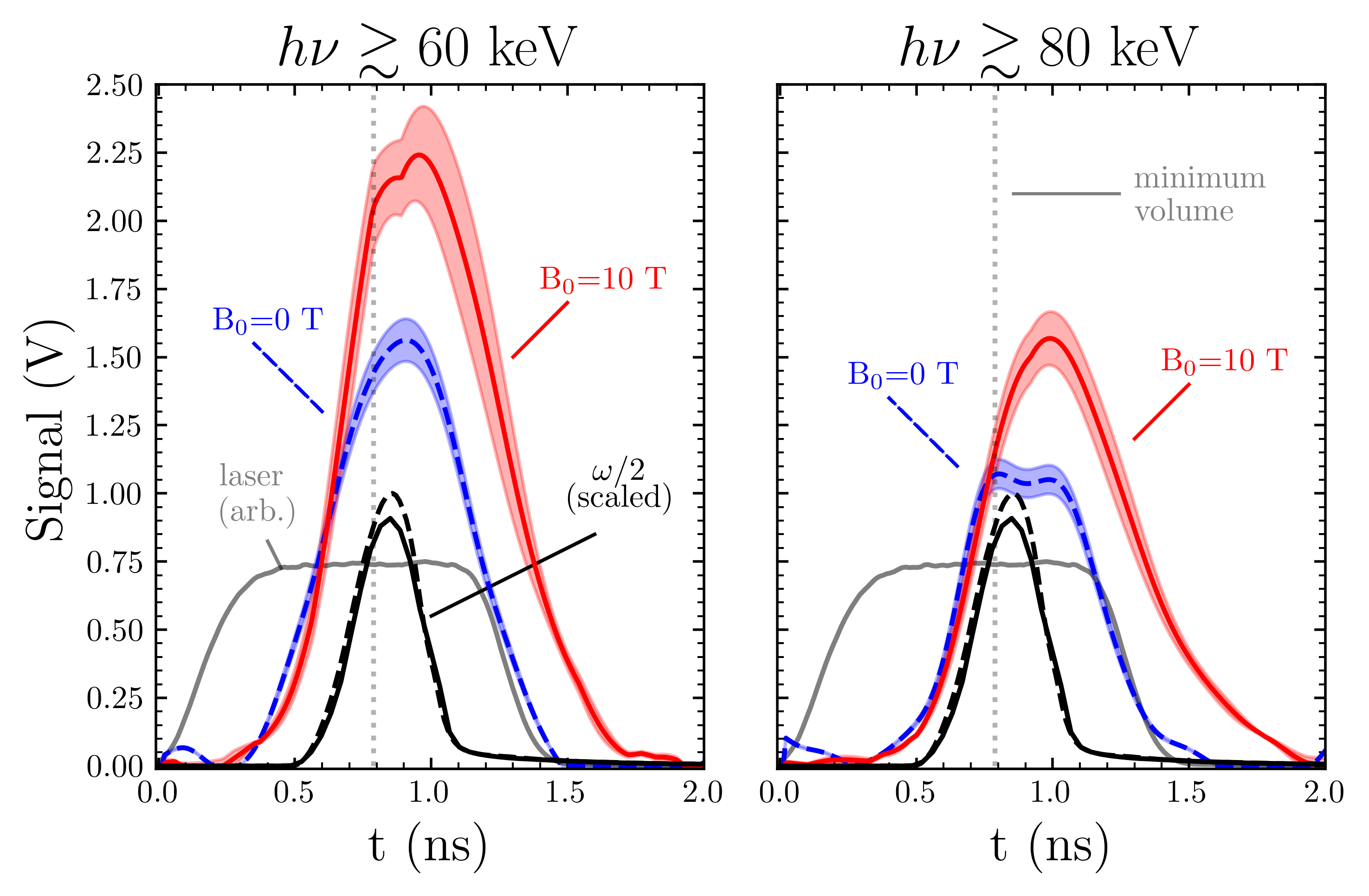"}
    \caption{Representative hard x-ray detector (HXRD) measurements for magnetized and unmagnetized implosions at OMEGA. Shaded regions are drawn to indicate the standard error in the measurement. These signals are associated with hot electrons generated by laser plasma interactions colliding with the shell and producing bremsstrahlung. The average $\omega/2$ signal from TPD scattered light in unmagnetized implosions is denoted by the dashed black curve and scaled to peak at unity; The average $\omega/2$ signal from magnetized implosions is denoted by the solid black curve, drawn on the same scale as the unmagnetized signal. The time of minimum volume is demarcated with the vertical, dotted, gray line. }
    \label{Fig:hxrd}
\end{figure}

Representative data from the hard x-ray detector (HXRD) are illustrated in Fig. \ref{Fig:hxrd}. The signal magnitudes are shown to systematically increase when the $10$ T magnetic field is applied. The ratio of the observed signal accumulated in the magnetized experiments compared to the unmagnetized experiments between 0 -- 1.5 \si{\nano\second} is $R_{60~\rm{kev}}\sim 1.52\pm 0.1$ for the $\gtrsim60 $ \si{\kilo\electronvolt} channel and is $R_{80~\rm{kev}} \sim 1.55\pm0.12$ for the $\gtrsim 80$ \si{\kilo\electronvolt} channel. 

\begin{figure}[p!]
    \vspace{1.0em}
    \includegraphics[width=\linewidth]{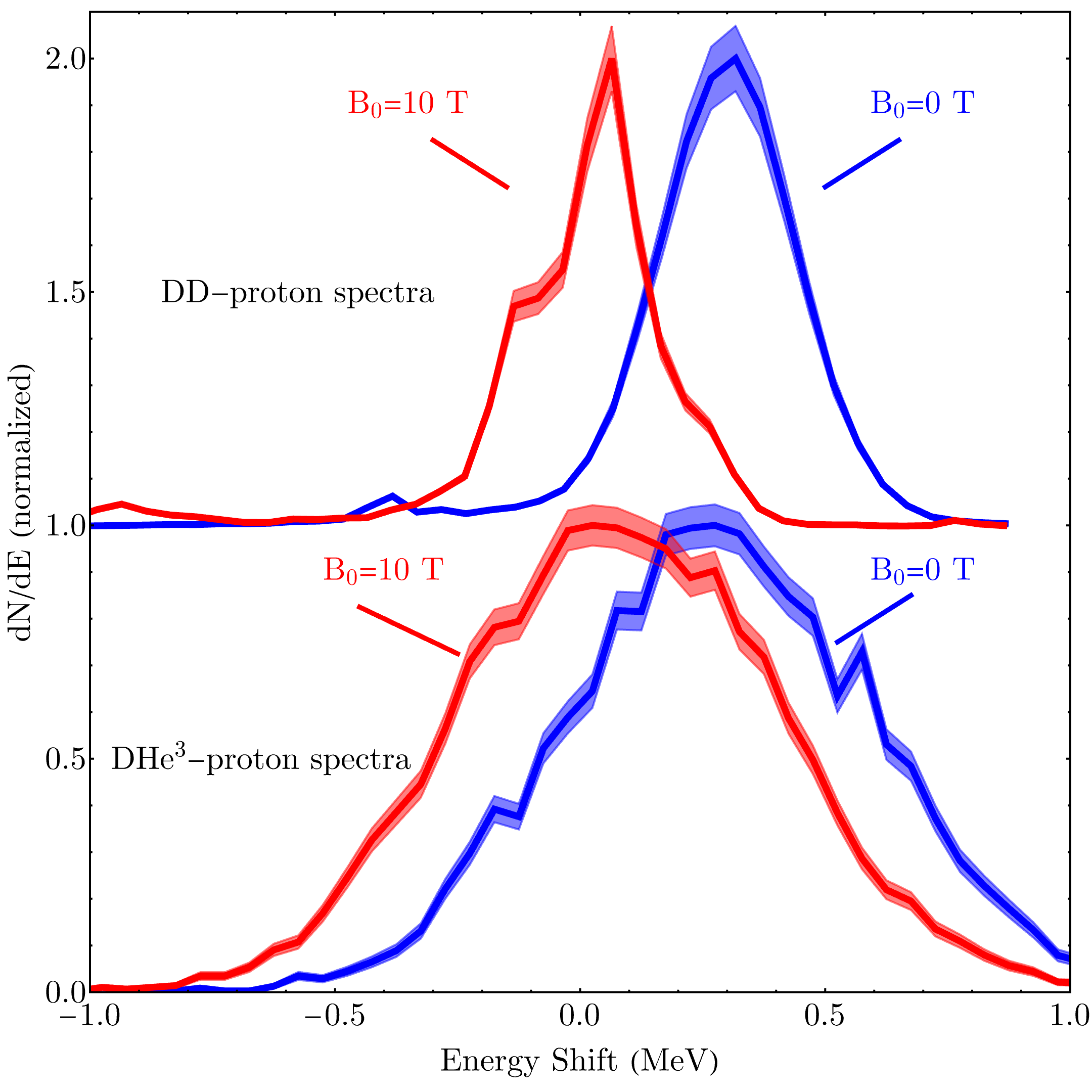}
    \caption{D$^3$He and DD proton spectra relative to the proton birth energy measured from  magnetized and unmagnetized experiments. In all cases, the median energy is seen to have shifted above the birth energy due to the capsule becoming positively charged during the laser drive. However, in magnetized implosions the capsule is evidently less charged.}
    \label{Fig:wrf:27kj}
\end{figure}

Examination of the proton spectra shown in Fig. \ref{Fig:wrf:27kj} reveals that the up-shift in energy is substantially smaller for magnetized implosions compared to unmagnetized implosions. Note that, since the plasma $\rho r$ is sufficiently small in these experiments, $\lesssim 5 $ \si{\milli\gram/\centi\meter^2}, variations in the measured particle energy introduced by the plasma stopping power may be neglected. Hence, the differences in median energy between unmagnetized to magnetized implosions may only be attributed to a reduction in the capsule charge. Similarly, it is observed that the shift in median energy in unmagnetized implosions is $\sim 300$ keV and in magnetized implosions is $\sim 50-100$ keV. These shifts are consistent for both DD ($\sim 3$ MeV) and D$^3$He protons ($\sim 14.7$ MeV). This further supports the conclusion that the shift is due to suppression of the capsule charge. 

\textit{Axial Magnetic Fields in Ablating Plasma Flows} The ablated shell plasma, extending from the critical surface to a distance of many capsule radii, is characterized by densities $~\sim 0.01 - 1~n_{\textrm{crit}}$ and temperatures $\sim 100~\textrm{\si{\electronvolt}} - 3~ \textrm{\si{\kilo\electronvolt}}$. The critical density is given by $n_{\textrm{crit}} \equiv 1.115\times 10^{21}/\lambda_\mu^2$ \si{\cm^{-3}} $= 9.1\times10^{21}$ \si{\cm^{-3}}, where $\lambda_\mu \approx 0.351$ \si{\micro\meter} is the laser wavelength -- the critical density provides the limit above which electromagnetic waves become evanescent in the plasma \cite{Michel_2023}. This plasma is unstable to many types of LPI in direct-drive ICF conditions that inhibit the coupling of laser energy to the capsule, reducing fusion yield and efficiency. Of principal concern here are instabilities that generate hot electrons: stimulated Raman scattering (SRS) and two-plasmon decay (TPD) \cite{rosenberg_2018}. TPD and SRS generally develop fairly late in the laser drive, when the density length scale of the ablated plasma $|n/\nabla n| \gtrsim 100 ~\mu$m -- in the experiments discussed here, measurements of scattered light from the full aperture backscatter station (FABS)\cite{regan_1999} diagnostic reveals TPD is predominant and peaks around $800$ ps after the laser drive begins (see Fig. \ref{Fig:hxrd}). These instabilities couple to the plasma in a three-wave process via an electron plasma wave (EPW) that accelerates electrons to energies in excess of $100 $ \si{\kilo\electronvolt} \cite{smalyuk_2008}. Some of these hot electrons strike the capsule surface and contribute to preheat. Prior work has examined the mechanics of hot electron acceleration/generation due to TPD in ICF relevant plasma conditions at the quarter-critical surface \cite{yan_2012} and applied these models to codes used in modeling direct-drive implosions \cite{cao_2022}. Other work has similarly examined the effect of an external magnetic field on LPI \cite{shi_2019,los_strozzi_2022,ghaffarioskooei2024} and hot electron transport in the corona/underdense plasmas of ICF implosions \cite{xxli_2023,xxli_2025}. Notably, transverse magnetic field configurations, where the magnetic field axis is orthogonal to the laser propagation axis, have been demonstrated to have a confining effect on the hot electron population in particle-in-cell (PIC) simulations, reducing the TPD instability saturation amplitude by increasing the rate of EPW damping \cite{xxli_2025}. However, as shown herein, the transverse magnetic field components are advected by the ablation flow over a timescale much shorter than the timescale for the development of TPD and SRS in direct-drive implosions. As such, large-amplitude magnetic fields in the corona transverse to the laser-propagation axis cannot be readily achieved during spherical compression of an initially axial field. Conversely, longitudinal magnetic field configurations, i.e. where the magnetic field is aligned with the laser propagation, have been demonstrated to have little impact on the generation of hot electrons at moderate magnetic field strengths, i.e., $B \lesssim 80$ T (\citealt{xxli_2025}; c.f. \cite{supp}).

\begin{figure}[p!]
    \centering
    \includegraphics[width=0.75\linewidth]{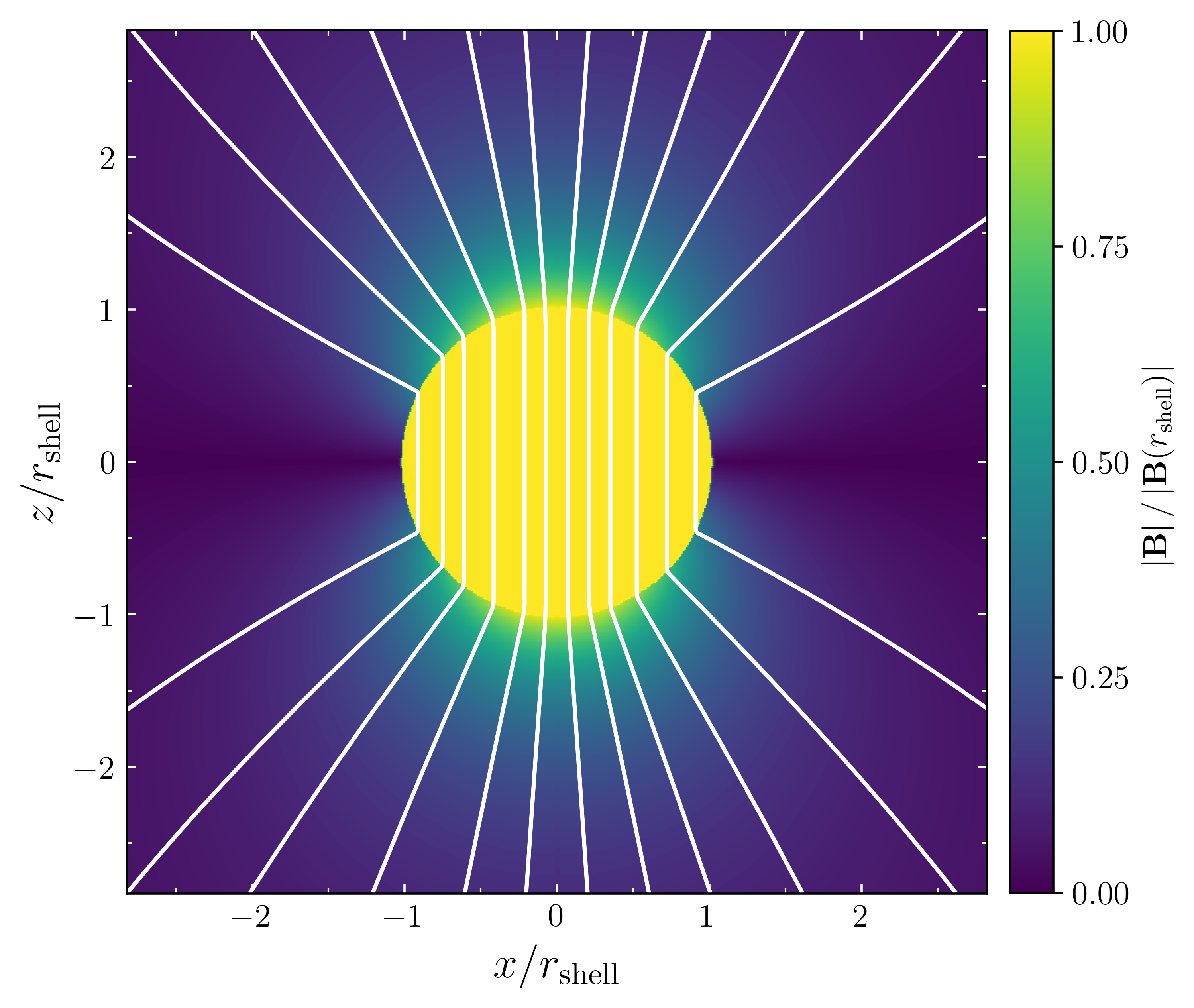}
    \caption{Numerical solution to ideal MHD model for the magnetic field evolution at $800$ ps for $u_{\rm{abl}} = 3 \times 10^8$ cm/s and $r_{\rm{shell}} = 5\times10^{-2}$ cm. The field is normal to the quarter-critical surface several $100$ picoseconds prior to the peak of TPD in our experiments. All field lines (white) are directed towards the top of the image.}
    \label{Fig:BFieldEvolution}
\end{figure}

In a magnetized direct-drive implosion, the capsule is immersed in an external, axial magnetic field $\textbf{B} = B_0 \hat{z}$. The laser drive rapidly establishes a radial, laminar ablation flow $u_{\rm{abl}} \hat{r} $ that terminates instantly at the ablation front near the critical surface. The kinetic energy associated with the flow is substantially larger than the magnetic energy, hence the back-reaction of the magnetic field on the plasma flow is regarded as insignificant. Similarly, magnetic field source terms, i.e. the Biermann battery, may be ignored since the self-generated fields seed transient, short length scale, fluctuations that are rapidly advected; the principal concern here is the form of the magnetic field field at length scales comparable to, or larger than, the capsule radius. Any variation in the azimuthal direction may be assumed to vanish, i.e. $\partial_\phi = 0$, due to the cylindrical symmetry of magnetized implosions. Lastly, since the magnetic Reynolds number is large for the ablating plasma flow, the magnetic diffusivity may be ignored. With these assumptions in mind, the magnetic field evolution is given by $\partial_t \textbf{B} = \nabla \times (\textbf{u}_{\textrm{adv}} \times \textbf{B})$
where $\textbf{u}_{\textrm{adv}}$ is the advection velocity of the plasma flow. Note that in the case of a spherical implosion, $\textbf{u}_{\textrm{adv}} = u_{\rm{abl}}\hat{r}$. In steady-state $\textbf{u}_{\rm{adv}}\times\textbf{B} = 0$ so the magnetic field must be radial, i.e., $\textbf{B} = B_r \hat{r}$. The functional form of $B_r$ is found from $\nabla \cdot B = 0$ and, using the boundary condition $B_r(r_{\rm{shell}}) = B_0 \cos\theta$ consistent with the applied field, the field is given by $B_r = B_0  r_{\rm{shell}}^2\cos\theta/r^2$ for $r > r_{\rm{shell}}$. The timescale for the field to reach the quasi steady-state may be estimated from the advection equation as $t_{\rm{field}} = r_{\rm{shell}}/u_{\rm{abl}}$. For typical ablation velocities $u_{\rm{abl}} = 3 \times 10^8 $ cm/s and shell radii $5 \times 10^{-2}$ cm, $t_{\rm{field}} \sim 200$ ps. During an implosion, the shell collapses and magnetic flux through the shell is well conserved. Therefore $B_r(t) = B_0(t) r_{\rm{shell}}(t)^2\cos\theta/r^2$ for $r > r_{\rm{shell}}(t)$ and $B_0(t) = B_0(0)(r_{\rm{shell}}(0)/r_{\rm{shell}}(t))^2$ for $r < r_{\rm{shell}}(t)$. Numerical integration of the magnetic field advection equation for $u_{\rm{abl}} = 3 \times 10^8$ cm/s and $r_{\rm{shell}} = 5 \times 10^{-2}$ cm is shown at $800$ ps in Fig \ref{Fig:BFieldEvolution}, illustrating that the magnetic field has reached the derived configuration before hot electron generation takes place in our experiments. 

The magnetic field at large radii must be considered to be the static applied field $\textbf{B} = B_0\hat{z}$ due to the coronal plasma's finite extent. The magnetic field amplitude $|\textbf{B}|$ then has a minimum in the corona and is maximized at the capsule surface and at the corona-vacuum interface, thereby trapping a population of electrons in a mirror-mode of the field \cite{Boyd_Sanderson_2003}. Of particular interest is the field amplitude near the quarter-critical surface (i.e., the locus where $n = n_{\rm{crit}}/4$) where hot electrons are generated by TPD. If the local minimum in $|B|$ is taken to be at the quarter-critical surface and the field assumed symmetric about that point, the timescale for hot electron transport in the corona is given by the period of an electron oscillation in the mirror-mode of the magnetic field, $t_{\rm{osc}} \sim4(r_{1/4}/v_{\perp})(1-\beta^2/2)$ where $\beta\equiv r_{\rm{shell}}/r_{1/4}$ is the ratio of the shell-radius to the quarter critical radius and $v_\perp$ is the electron's transverse velocity \cite{supp}. For hot electrons, $v_\perp \gtrsim 10^{10}$ \si{\cm/\second} (equivalent to $E_{\rm{kin}} \gtrsim 30$ keV) it is inferred from 1D radiation hydrodynamic simulations that, at the time the TPD instability peaks, $\beta \sim 1/4$ and $r_{1/4} = 0.1$ \si{\centi\meter} giving a timescale of $t_{\rm{osc}} \sim 50$ \si{\pico\second}. Note that the ordering of the timescales $t_{\rm{field}} \ll 800~$\si{\pico\second} allows us to consider the magnetic field to have already reached the quasi-steady state when the TPD amplitude is maximized. The timescale for laser-driven instabilities in direct-drive ICF to saturate is $t_{\rm{LPI}} \lesssim 10$ \si{\pico\second} $ \ll 800~$ \si{\pico\second}. Therefore, any LPI  may be assumed to take place in a radial magnetic field $\textbf{B} = B_r \hat{r}$ and instantly saturate on the timescales relevant to hot electron transport. 

\textit{Discussion} In direct-drive ICF implosions, the capsule charge may be attributed to the ejection of a population of hot electrons from the coronal plasma produced via TPD. To explain the reduction in charge one must now consider the single-particle motion of these hot electrons in the quasi-steady state magnetic field. Analysis of the hot electrons' motion in this magnetic field, and in the absence of a magnetic field, reveals that the reduction in capsule charge is directly related to the increase in hot electron preheat. Notably, the possibility that the data are explained by modification of TPD instability and hot electron generation is ruled out by PIC simulations\cite{supp}.  

For the case of unmagnetized electron motion in a non-neutral collisionless plasma, capsule charging will initially occur for electrons with sufficient angular momentum and energy -- the conditions on the electron pitch-angle and energy can be derived exactly for a non-neutral collisionless plasma \cite{supp} but are well approximated by the conditions $v_\perp/v_\parallel \gtrsim \beta$ and $v_\perp^2+v_\parallel^2 \gtrsim 2 \beta\tilde{U}$ where $\tilde{U}$ is the potential energy of an electron at the capsule shell in units of the electron rest mass $m_ec^2$ (see supplemental material \cite{supp}). The former condition can be understood as a statement of sufficient angular-momentum to avoid striking the capsule and the latter can be understood as the condition for an electron to be energetically unbound from the capsule. 

Now consider the motion of an electron in a non-neutral plasma immersed in the quasi steady-state magnetic field. The electron has a magnetic moment associated with its gyromotion about the field line $\mu \equiv m_e v_\perp^2/(2B)$. This moment will be well conserved when the corona crossing frequency is less than or about the gyro-frequency. The typical corona crossing frequency is $\omega_{\rm{osc}} = 2\pi/t_{\rm{osc}} \sim 1.2 \times 10^{11}$ rad/s and the associated gyro period of an electron at the quarter critical surface for an implosion with external field strength $10 $ \si{\tesla}, convergence ratio of $4$, and $\beta = 1/4$ is $\omega_{c\bar{e}}\sim 2\cos\theta \times 10^{12}$ rad/s where, $\theta$ is the angle between the initial magnetic field axis and the field line about which the electron is gyrating. One should expect then that the magnetic moment $\mu$ is well conserved everywhere in the corona except for a very narrow region around the equator. Since the magnetic field is convex -- i.e. is maximized at $r_{\rm{shell}}$ and at the edge of the coronal plasma -- there must be a mirror-like trapping criterion \cite{Boyd_Sanderson_2003}. Conservation of the electron magnetic moment and electron energy in the derived field give $v_\parallel^2 +v_\perp^2(1-1/\beta^2) < 2\tilde{U}(\beta-1)$ as the condition for trapping in a mirror-mode of the magnetic field where $\tilde{U}$ is the capsule potential energy in units of $m_ec^2$. This condition must also be supplemented by a condition on the particle energy when $\tilde{U} \neq 0$ \cite{supp}. That is, electrons that collide with the capsule and produce bremsstrahlung are those that are born with pitch-angles $\alpha \equiv \arctan v_\perp/v_\parallel \lesssim \beta$, where $v_\parallel$ is the component of the electron velocity direction towards the capsule surface and $v_\perp$ is in the orthogonal direction. Similarly, electrons born at very large pitch angle $\pi - \alpha \lesssim \beta$, i.e., the nearly backward direction, will escape the corona and contribute to capsule charging. These conditions define the `x-ray cone' as the locus of electron velocity-vectors that result in trajectories that intersect the capsule and the `charging cone' as the locus of electron velocity-vectors that result in trajectories that escape the capsule potential. In magnetized implosions, electrons with velocity vectors that point between the charging cone and x-ray cone are confined in a mirror-mode of magnetic-field and will scatter into the cones. Critically, electrons in the mirror-region of a magnetized implosion would otherwise escape and contribute to charging in an unmagnetized implosion \cite{supp}.  

The timescale for scattering of a hot electron $v_{\bar{e}}\sim 10^{10}$ cm/s on the cool, thermal, population of electrons in the corona is $t_{\rm{col}} \sim 40 - 70$ ps \cite{aBers_book}. This timescale is comparable to $t_{\rm{osc}}$ for the electrons mirroring in the magnetic field and therefore the increase in hard x-ray signal in magnetized implosions may be attributed to the population of hot-electrons initially confined in the mirror-region of the magnetic field scattering onto the capsule surface. The difference in capsule charge is naturally explained as well since the confined electrons would otherwise escape the corona and contribute to charging \cite{supp}.

\begin{figure}[p!]
    \centering
    \includegraphics[width=\linewidth]{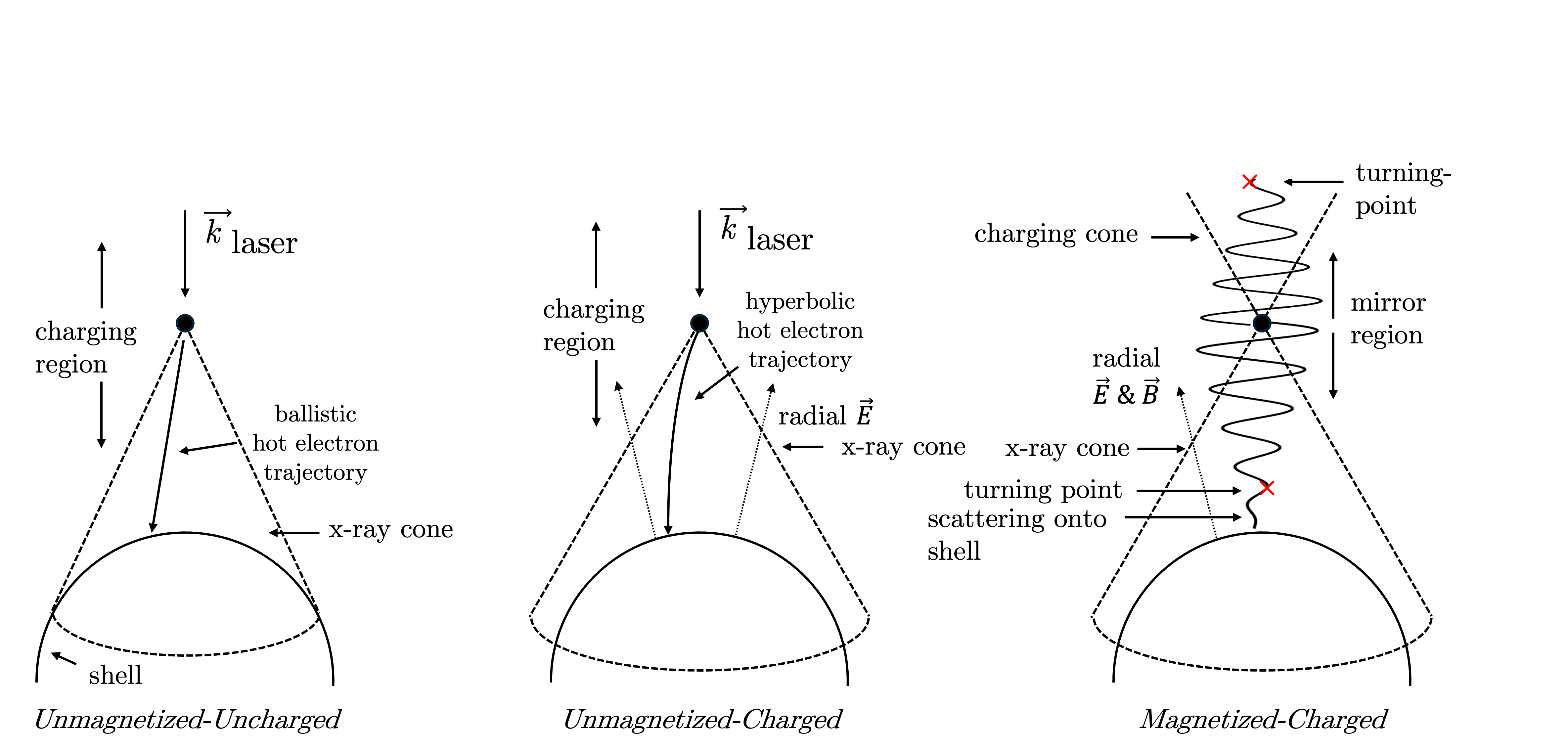}
    \caption{Diagrams illustrating the hot electron trajectories in various implosions. In all cases, hot electrons with trajectories inside the x-ray cone will collide with the capsule and generate preheat. In unmagnetized implosions, hot electrons with trajectories outside the x-ray cone will contribute to capsule charging. Conversely, in magnetized implosions, due to confinement by the magnetic field and pitch-angle scattering, electrons with trajectories in the mirror-region will strike the capsule surface and contribute to preheat.}
    \label{Fig:lossCone}
\end{figure}

Illustrations of the various possibilities for hot electron trajectories are summarized in Fig. \ref{Fig:lossCone} which show the charging/x-ray/mirror regions for hot electrons generated at the capsule pole in implosions with and without magnetic fields. Hot electrons with velocity vectors inside the x-ray cone collide with the capsule regardless if the implosion is magnetized or unmagnetized. Electrons with velocity vectors outside the x-ray cone, i.e. in the charging region, escape and lead to capsule charging in unmagnetized implosions. In magnetized implosions, the charging-region is partially replaced with the mirror-region, what remains is denoted as the charging cone. Electrons in the mirror-region are confined in the magnetic field until they pitch-angle scatter into either the x-ray or charging cones.

After the TPD instability saturates, the hot electron velocity distribution at the quarter critical surface may be approximated as a two-temperature maxwellian $f \propto \exp(-m_e v_\parallel^2/2T_{\rm{hot}} - m_ev_\perp^2/2T_{\rm{cold}})$, where the electrons' characteristic temperature in the direction towards the capsule surface is $T_{\rm{hot}}$ and in the other directions the characteristic temperature is $T_{\rm{cold}}$. Furthermore, the hard x-ray signal in an unmagnetized implosion may be assumed to be proportional to the number of hot electrons in the x-ray cone with energy greater than the detector cut-off, 60 keV or 80 keV. Similarly, the excess hard x-ray signal in magnetized implosions may be assumed to be proportional to the number of electrons confined in the mirror-mode of the magnetic field. With these assumptions the ratio of magnetized preheat to unmagnetized preheat may be approximated as $R \sim 1 + I_1/I_2$, where, $I_1$ is the integral of the hot-electron distribution excluding contributions from the x-ray cone and electrons with energy less than the detector cut-off. Similarly, $I_2$ is the integral of the hot-electron distribution over the x-ray cone excluding electrons with energy less than the detector cut-off. Using the formulations of the x-ray cone in \cite{supp} and assuming that $T_{\rm{cold}} = 3$ keV and $\beta \sim 1/4$ -- inferred from 1D Hyades \cite{hyades} radiation hydrodynamic simulations -- and $T_{\rm{hot}} = 90$ keV as in Ref. \cite{froula_2012}, evaluation of the preheat ratio gives $R_{60~\rm{kev}} \sim 1.6$ and $R_{80~\rm{kev}} \sim 1.51$ similar to the measured values. The capsule charge ratio may be estimated similarly.

\textit{Summary} Experimental evidence has been presented that demonstrates that hot electron preheat is increased in magnetized direct-drive implosions by a factor $ 1.5\pm 0.1$. The hot electron population produced via LPI is effectively confined by the application of an initially $10$ \si{\tesla} magnetic field to an ICF implosion in a mirror-mode of the magnetic field and efficiently pitch-angle scatters onto the capsule surface; this model self-consistently explains the observed increase in hard x-ray emission and the decrease in capsule charge. 

Prior work examining the effect of magnetization on LPI in direct-drive relevant conditions has noted that the presence of a transverse magnetic field suppresses the TPD instability growth-rate and leads to local trapping of hot electrons at the quarter critical surface thereby mitigating capsule preheat \cite{xxli_2025}. However, the work here illustrates that transverse components of the magnetic field are rapidly advected in the ablating plasma flow prior to the development/saturation of the electron-generating LPI. The field reaches a quasi steady-state that is radial in spherical implosions and does not appreciably affect the LPI development or thresholds for currently experimentally realizable magnetic field strengths $\sim 10$ T. Consequently, mitigating LPI in direct-drive implosions, e.g., via the use of broadband lasers, is of increased importance in magnetized implosions, not less as has been previously suggested. 

Previous experiments investigating the impacts of an applied magnetic field on laser-driven implosions have been limited to a reduced number of laser beams with access to the capsule waist inhibited by the coils used to generate the magnetic field, necessitating pointing the beams in a polar-drive configuration \cite{marshall_2006}. Experiments using this drive configuration noted enhancement in the apparent ion-temperature and yield \cite{chang_2011} and restricted electron heat-conduction across the magnetic field \cite{bose_2022}. Notably, these previous experiments were necessarily performed at lower laser intensities than those reported here and electron preheat is not expected to have contributed to the dynamics/fuel conditions in those experiments. This motivates a new experimental campaign to investigate the effect of enhanced electron preheat on fuel compressibility and fusion yield in directly-driven magnetized implosions.

Additionally, the results presented here have significant implications for radiation-magnetohydrodynamic codes used to model ICF implosions since existing codes do not self-consistently treat the effects of external magnetic fields on hot electron generation, trapping, and the increased preheat. An inability to account accurately for the effect of hot electron preheat will lead to systematic inconsistencies in the modeling of magnetized implosion experiments. 
This effect may be accounted for in simulations in an ad-hoc manner by modifying existing electron-preheat treatments with a multiplier determined from the preheat ratio discussed here and should be explored in the future.  

\section*{Acknowledgments}
This work was supported in part by the U.S. Department of Energy NNSA MIT Center of Excellence under Contract No. DE-NA0003868. M. Cufari is supported by NNSA SSGF under Grant No. DE-NA0003960. The experiment was conducted at the Omega Laser Facility with the beam time through the National Laser Users’ Facility (NLUF) (or the Laboratory Basic Science) under the auspices of the U.S. DOE/NNSA by the University of Rochester’s Laboratory for Laser Energetics under Contract DE-NA0003856. This material is based upon work supported by the Department of Energy [National Nuclear Security Administration] University of Rochester “National Inertial Confinement Fusion Program” under Award Number(s) DE-NA0004144.

Disclaimer: This report was prepared as an account of work sponsored by an agency of the United States Government. Neither the United States Government nor any agency thereof, nor any of their employees, makes any warranty, express or implied, or assumes any legal liability or responsibility for the accuracy, completeness, or usefulness of any information, apparatus, product, or process disclosed, or represents that its use would not infringe privately owned rights. Reference herein to any specific commercial product, process, or service by trade name, trademark, manufacturer, or otherwise does not necessarily constitute or imply its endorsement, recommendation, or favoring by the United States Government or any agency thereof. The views and opinions of authors expressed herein do not necessarily state or reflect those of the United States Government or any agency thereof.


\begin{figure}[p!]
    \centering
    \includegraphics[width=\linewidth]{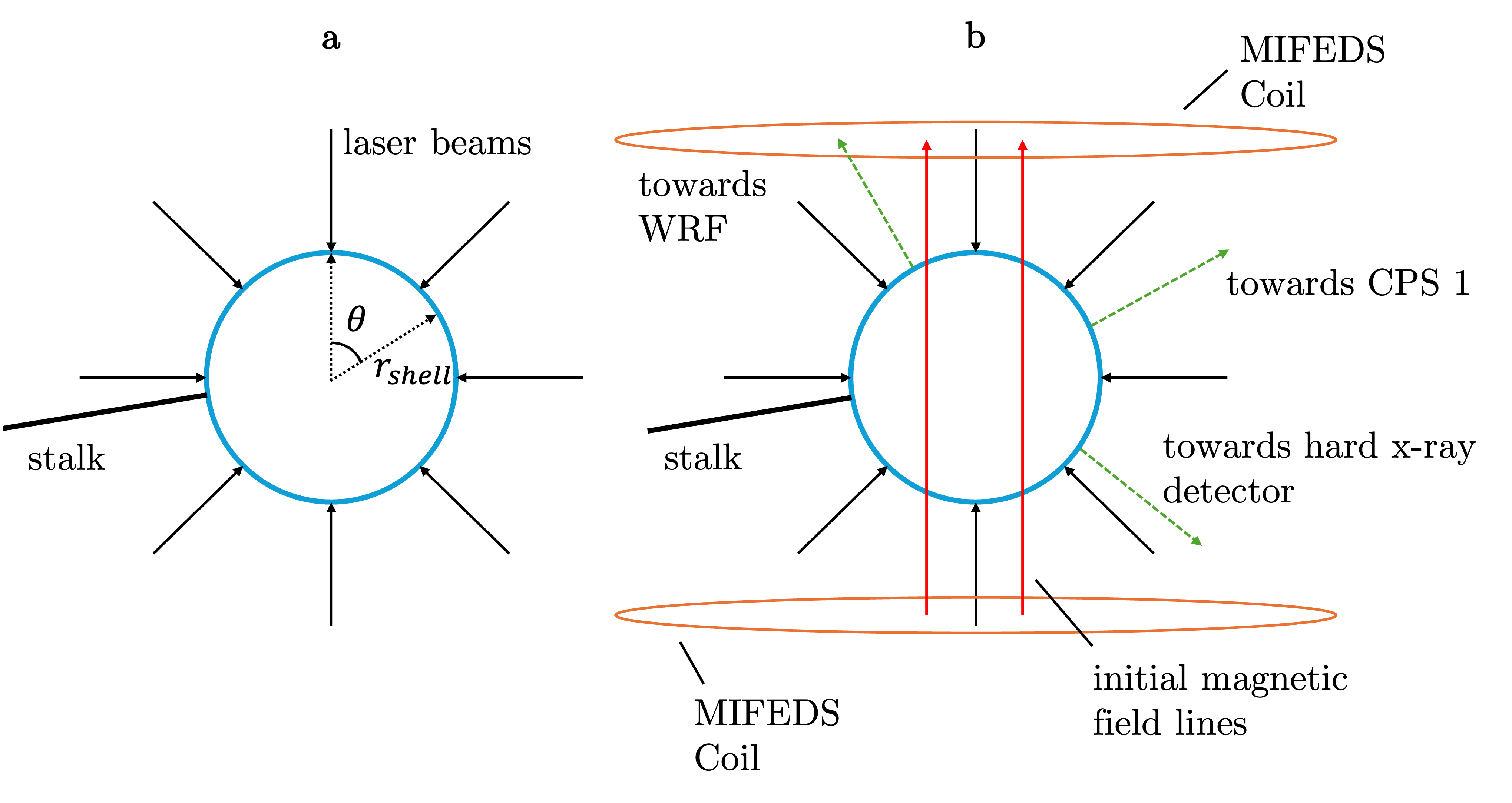}
    \caption{Experimental diagrams from a) unmagnetized and b) magnetized implosions. MIFEDS coils are not drawn to scale. Lines of sight from the diagnostics indicated are unobstructed by the coils.}
    \label{Fig:expDiagram}
\end{figure}

\pagebreak

\widetext
\begin{center}
\textbf{\large Supplemental information for Enhanced Hot Electron Preheat Observed  in Magnetized Laser Direct-Drive Implosions}
\end{center}
\setcounter{equation}{0}
\setcounter{figure}{0}
\setcounter{table}{0}
\setcounter{page}{1}
\makeatletter
\renewcommand{\theequation}{S\arabic{equation}}
\renewcommand{\thefigure}{S\arabic{figure}}

\section{Electron-bounce frequency in the quasi steady-state magnetic field}
For completeness, a derivation of the electron bounce-frequency in the adiabatic approximation is given in this supplemental material. We analyze the result in the limit of large $r_{1/4}/r_{\rm{shell}}$, where $r_{1/4}$ denotes the radius of quarter-critical density surface and $r_{\rm{shell}}$ denotes the imploding shell radius. For convenience, we define $r_{\rm{shell}}/r_{1/4} \equiv \beta $ to have $\beta \ll 1$. The adiabatic assumption of a quasi steady-state field is justified \textit{a posteriori}. 

The radial component of an electron's equation of motion is, in the adiabatic approximation, given by
\begin{equation}
         \dot{v}_r = \frac{-\mu}{m_e} (\hat{r}\cdot\nabla) B_r
\end{equation}
where $\mu = m_e v_\perp^2/(2B)$ is the electron's magnetic moment. Here, the radial direction is along the normal of the imploding shell surface. A quasi steady-state, radial magnetic field of $B_r = B_0 r_{\rm{shell}}^2 \cos\theta\hat{r}/r^2$, valid for  $r > r_{\rm{shell}}$, was given in the paper. Inserting this expression into the equation of motion yields
\begin{equation}
         \dot{v}_r =\frac{2B_0 \cos\theta \mu}{m_e}\frac{r_{\rm{shell}}^2}{r^3}.
\end{equation}
Using the relation $\dot{v_r} = v_r \partial v_r/\partial r$ and integration bounds from the imploding shell $r_{\rm{shell}}$ to an arbitrary point in space $r$ away from the shell surface, one obtains
\begin{equation}
\frac{1}{2}v_r^2 = \frac{B_0\cos\theta \mu}{m_e} \left(1-\frac{r_{\rm{shell}}^2}{r^2}\right).
\end{equation}
Using $v_r^2 = (dr/dt)^2$ and that the transit from the shell to the quarter critical surface is approximately one-quarter of the total distance traversed in a full oscillation period, the electron oscillation period is
\begin{equation}
 	t_{\rm{osc}} =  4 \sqrt{\frac{m_e}{2\mu B_0\cos\theta}} \int_{r_{\rm{shell}}}^{r_{1/4}} \frac{dr}{\sqrt{1-r_{\rm{shell}}^2/r^2}}, 
\end{equation}
where the factor in front of the integral can be identified as $4/v_\perp$. Evaluating this integral gives
\begin{equation}
t_{\rm{osc}} =  4 \frac{r_{1/4}}{v_\perp} \sqrt{1-\beta^2} ,
\end{equation} 
For hot electrons, $v_\perp \sim 10^{10}$ cm/s and taking $r_{\rm{1/4}} \sim  10^{-1}$ cm,  the oscillation timescale is $t_{\rm{osc}} \sim 50$ ps. Compared to the magnetic field evolution timescale, $t_{\rm{field}}\equiv r_{\rm{shell}}/v_{\rm{abl}}= 500 \times 10^{-4} \rm{cm}/3 \times 10^8 \rm{cm/s} \sim 200$ ps, the oscillation time is much shorter. The evolution of the magnetic field is thus unimportant to the electron motion in the field and the adiabatic approximation is well motivated.

\section{Ordering of the relevant time-scales}
As shown in the preceding section, a hot electron bounces in the coronal magnetic field on a time-scale of $t_{\rm{osc}} \sim 50$ ps. It has also been argued that the magnetic field evolves on a slower timescale of $t_{\rm{field}} \sim 200$ ps. The experimental time-scale of $t_{\rm{hydro}} = 1$ ns is fixed by the total duration of the laser-drive and limits the physics of interest to shorter timescales. The two-plasmon decay (TPD)  instability is observed to peak near the end of the laser pulse $t \sim 800$ ps in experiments and may be considered to reach a saturated state within $t_{\rm{LPI}} \sim 10$ ps of the onset of the instability. This is the shortest time-scale relevant to the results and discussion in the paper. With this in mind, the ordering of the timescales are
\begin{equation}
	t_{\rm{LPI}} \ll t_{\rm{osc}} \ll t_{\rm{field}} < t_{\rm{hydro}}
\end{equation}
where the last relation is understood to hold only sufficiently close to the imploding shell, i.e, $r < u_{\rm{abl}}t_{\rm{hydro}}  \sim 10^{-1}$ cm. 

\section{The electrons that contribute to fuel preheat}
\begin{figure}
    \includegraphics[width=\linewidth]{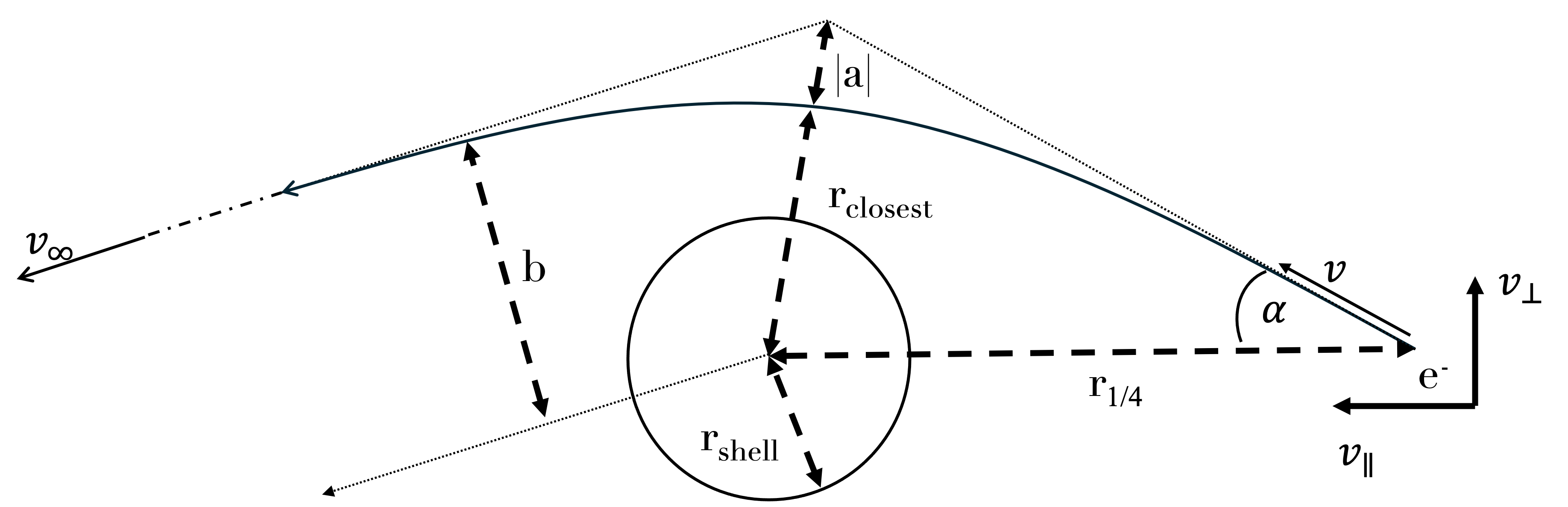}
    \caption{Illustration of a hot electron's hyperbolic trajectory about the charged shell. Charging occurs for $r_{\rm{closest}} > r_{\rm{shell}}$}
    \label{fig:electronTrajectoryCartoon}
\end{figure}

Here we consider the problem of determining which hot-electrons generated from TPD strike the shell and contribute to preheat when there may be a global electric field and strong external magnetic field. We approach this problem by considering the conditions on the electron pitch-angle $\alpha = \arctan v_\perp/v_\parallel$, (Fig. \ref{fig:electronTrajectoryCartoon}), which lead to shell charging during an implosion in the absence of a magnetic field -- since electrons may only either strike the capsule or contribute to shell charging, this is condition is sufficient to determine which electrons will strike the shell. First, note that the population of hot electrons ought to be collisionless with the background coronal plasma since $v_{\bar{e}}/\nu_{\bar{e}e} \gg r_{\rm{shell}}$, where $v_{\bar{e}}$ is the typical velocity of hot electrons $v_e \sim 10^{10}$ cm/s and $\nu_{\bar{e}{e}}$ is the collision frequency of hot electrons with $T_{\bar{e}} = 90$ keV on coronal electrons with $T_e = 3$ keV. Second, consider the imploding shell to be a perfect conductor with charge $eQ$, established by the ejection of $Q$ electrons from the plasma during the laser drive. The binding energy of an electron located at the shell surface is $U = eV = e^2Q/4\pi\epsilon_0 r_{\rm{shell}}$. For $v^2 < 2\beta U /m_e$, where $\beta \equiv r_{\rm{shell}}/r_{1/4}$, the electron is energetically bound to the shell and cannot contribute to charging. Hereafter, physical quantities are scaled such that velocities are in units of the vacuum speed of light $c$ and energies in units of the electron rest-mass $m_e c^2$; scaled quantities are denoted with a tilde, e.g., the escape speed is $\tilde{v}_{\rm{esc}}^2 \equiv 2 \beta \tilde{U}$. 

Hot electrons with velocities greater than the escape speed are on hyperbolic trajectories around the shell as illustrated in Fig. \ref{fig:electronTrajectoryCartoon}. Since these electrons are produced at the quarter-critical surface, the impact parameter of the electron trajectory is $b = r_{1/4}\tilde{v}_\perp/\tilde{v}_{\infty}$ where $\tilde{v}_{\infty}^2 = \tilde{v}^2 - \tilde{v}_{\rm{esc}}^2$ is the electron velocity infinitely far from the shell. The electron trajectory semi-major axis is $a = - r_{\rm{shell}} (\tilde{U}/ \tilde{v}_{\infty}^{2})$. The electron distance of closest approach is $r_{\rm{closest}} = \sqrt{a^2+b^2}+a$. For an electron to escape the shell potential and increase the shell-charge, $r_{\rm{closest}} > r_{\rm{shell}}$. Therefore, $\sqrt{a^2+b^2} + a > r_{\rm{shell}}$ gives the condition on $v_\perp$ and $v_\parallel$ that leads to charging. Inserting the definitions of $a$ and $b$ into this condition gives
\begin{equation}\
	r_{\rm{shell}} < \sqrt{\frac{\tilde{U}^{2} r_{\rm{shell}}^2}{\tilde{v}_{\infty}^4} + \frac{\tilde{v}_{\perp}^{2} r_{1/4}^{2}}{\tilde{v}_{\infty}^2} } - \frac{\tilde{U} r_{\rm{shell}}}{\tilde{v}_{\infty}^2}.
\end{equation} 
Eliminating $r_{\rm{shell}}$ and multiplying the equation by $\tilde{v}_{\infty}^{2}$ gives
\begin{equation}\label{eq:1}
\tilde{v}_{\infty}^2 + \tilde{U} < \left(\tilde{U}^2 + \frac{\tilde{v}_{\infty}^{2} \tilde{v}_{\perp}^2}{\beta^2}\right)^{1/2}.
\end{equation}
It has been established that the electrons are on hyperbolic trajectories and $\tilde{v}_{\infty} > 0$. It will be shown momentarily that this restriction reduces the problem from quartic order in $\tilde{v}_{\parallel}$ and $\tilde{v}_{\perp}$ to quadratic. Using $\tilde{v}_{\infty}^2 = \tilde{v}_{\parallel}^2+\tilde{v}_{\perp}^2 - 2\beta \tilde{U}$  and squaring yields
\begin{equation}\label{eq:vperpparCond}
	(\tilde{v}_{\parallel}^2 + \tilde{v}_{\perp}^2 + \tilde{U}(1-2\beta))^2 < \tilde{U}^2 + \frac{\tilde{v}_{\perp}^2}{\beta^2}(\tilde{v}_{\perp}^2 + \tilde{v}_{\parallel}^2 - 2 \tilde{U}\beta). 
\end{equation}
This equation may be factorized into a product of polynomials 
\begin{equation}
	(\tilde{v}_{\parallel}^2 + \tilde{v}_{\perp}^2 - 2\beta \tilde{U})\left(\tilde{v}_{\parallel}^2 + \tilde{v}_{\perp}^2\left(1-\frac{1}{\beta^2}\right) + 2 \tilde{U}(1-\beta)\right) < 0.
\end{equation}
The first is identified as a condition for the electrons to escape the potential, i.e., $v_{\infty} > 0$ hence, this term is always positive. Therefore the above equation can be split into two quadratic expressions that are required to be simultaneously satisfied:
\begin{align} 
	\tilde{v}_{\parallel}^2 + \tilde{v}_{\perp}^2\left(1-\frac{1}{\beta^2}\right) &< 2 \tilde{U}(\beta-1), \\
	\tilde{v}_{\parallel}^2 + \tilde{v}_{\perp}^2 > 2\beta \tilde{U}.
\end{align}
The approximate condition given in the paper $v_\perp/v_\parallel > \beta $ for electron charging is obtained in the limit that $\tilde{U} = 0$ and by using the approximation $\left(\beta^{-2}-1\right)^{-1/2} \sim \beta$, valid in the limit $\beta \ll 1$. The condition for an electron to strike the capsule is obtained by reversing the direction of the inequality in (11).

Now, the condition for electron confinement in the presence the quasi steady-state magnetic field is considered. The magnetic field has the form, $\vec{B} = r_{\rm{shell}}^2 \cos\theta \hat{r} /r^2$ (as established in the paper). An electrostatic potential due to the imploding shell becoming positively charged during the laser drive is also present. The adiabatic approximation yields $\mu \equiv m_e v_\perp^2 /2B$ as a constant of electron motion in the field. Since the magnetic field cannot exert work, the condition for an electron to be energetically unbound in the shell potential is not affected by the presence of the magnetic field and remains the same as in the absence of the magnetic field, $\tilde{v}_\perp^2 + \tilde{v}_\parallel^2 >  2\beta \tilde{U}$. Conservation of the magnetic moment implies that
\begin{equation}
\frac{{B_{1/4} }}{B_{\rm{shell}}}= \frac{\tilde{v}_{1/4\perp}^{2}}{\tilde{v}_{\rm{shell}\perp}^2},
\end{equation}
where subscript `1/4' or `shell' indicates the radial location at which the variable is evaluated. For a particle with turning point coincident with the imploding shell, conservation of energy implies that 
\begin{equation}
\tilde{v}_{1/4\parallel}^2 +\tilde{v}_{1/4\perp}^2 - 2\beta \tilde{U} = \tilde{v}_{\rm{shell}\perp}^2 - 2\tilde{U},
\end{equation}
where $v_{\rm{shell}\parallel}^2 = 0$ is used on the right-hand side. Combining this equation and the conservation of magnetic moment, noting $B_{1/4}/B_{\rm{shell}}  = \beta^2$, and enforcing that the turning point for electrons oscillating in the corona simply be at larger $r$ than $r_{\rm{shell}}$ we find that electrons are confined for
\begin{equation}
	 \tilde{v}_{\parallel}^2 + \tilde{v}_{\perp}^2 \left(1-\frac{1}{\beta^2}\right) < 2\tilde{U}(\beta-1).
\end{equation}
When supplemented with the energy condition above, this confinement condition for the magnetic field is observed to be identical to the charging condition in the absence of the magnetic field. Therefore, it is the electrons that are confined by the field in magnetized implosions -- and would otherwise charge the shell in unmagnetized implosions -- that strike the shell and contribute to the enhancement of fuel preheat. A small number of electrons in the charging-cone, $\pi-\beta < \alpha < \pi + \beta$, contribute to residual charging of the shell. 

\section{\label{sec:simulations} Simulations}
Two PIC simulations were performed using the OSIRIS code \citep{OSIRIS} to investigate how aligning a magnetic field along the laser propagation axis affects the hot electron generation. Both simulations were performed in a cartesian box with dimensions $34$ \si{\micro\meter}$\times~11.3$ \si{\micro\meter} and utilized a laser intensity $a_0 = 0.0073 = 5.7\times10^{14}$ W/cm$^2$.  The laser had a wavelength of $0.355$ \si{\micro\meter} and was treated as a gaussian envelope, with full-width-half max of $\Delta R = 1000~ c/\omega_0$ where $\omega_0$ is the laser frequency. The x-axis in the PIC simulations may be interpreted as being aligned with the radial direction and the y-axis is interpreted as being aligned azimuthally. As such, the laser propagates parallel to the x-axis. The laser used in this simulation was polarized parallel to the y-axis and launched from the $x = 0$ boundary. The laser had a rise time of $300/\omega_0$ after which it saturated at its maximum value for the duration of the simulations. One simulation was performed in which a static, external magnetic field was imposed pointing parallel to the x-axis, and the other simulation imposed no external magnetic field. The magnetic field strength was $10 $~\si{\tesla} and corresponds to a cyclotron period $\omega_c/\omega_0 = 0.00033$. A density gradient from $n_e = 0.21~n_{\textrm{crit}}$ -- $0.27~n_\textrm{crit}$ was imposed parallel to the x-axis. The scale length of the density gradient was $L = n/\nabla n = 139.3$ \si{\micro\meter} at the quarter critical surface. The initial electron temperature was $3$ \si{\kilo\electronvolt}, the ion temperature was $1.5$ \si{\kilo\electronvolt}, and the TPD instability parameter for these conditions is $\eta = L_\mu\lambda_\mu I_{14}/(81.86~T_{kev}) = 1.16$ \citep{simon_1983}. The ions have a reciprocal charge-to-mass ratio of $3000$. Periodic boundaries were assumed for the y-boundaries and thermal reflective boundaries at the initial temperature were used for the x-boundaries. The electric and magnetic fields used open boundary conditions at the x-boundaries and periodicity was assumed for the y-boundaries. The simulations were performed out to $8.3$ \si{\pico\second}. 

\begin{figure}[h]
     \centering
     \includegraphics[width=\linewidth]{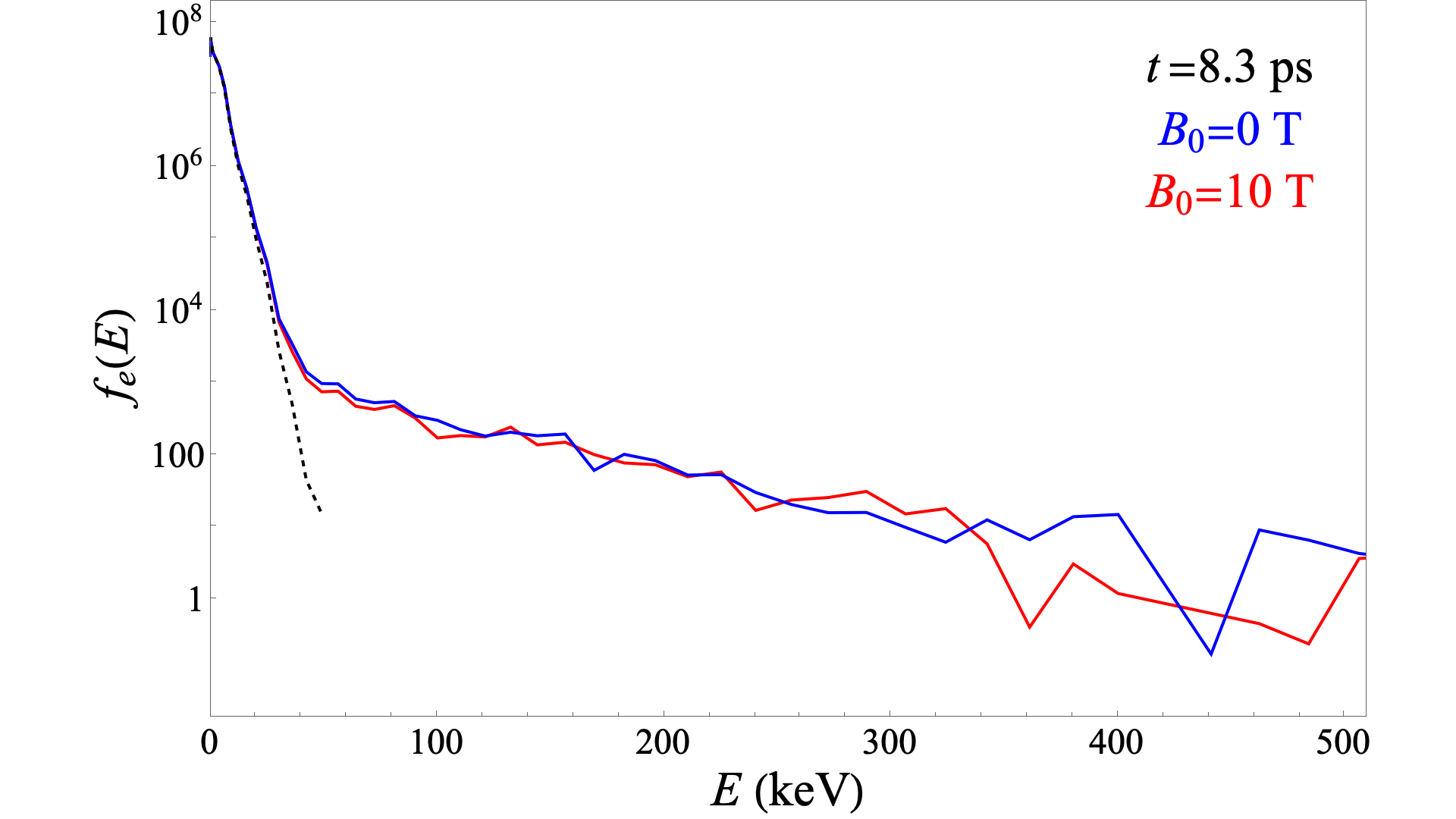}
     \caption{Electron energy distribution functions at $8.3$ \si{\pico\second} from magnetized (red) and unmagnetized (blue) PIC simulations. The dashed black line is drawn to indicate the initially thermal distribution function at $t=0$}\label{fig:electronEnergySpectrum}
\end{figure}

\begin{figure}
\centering
\includegraphics[width=\linewidth]{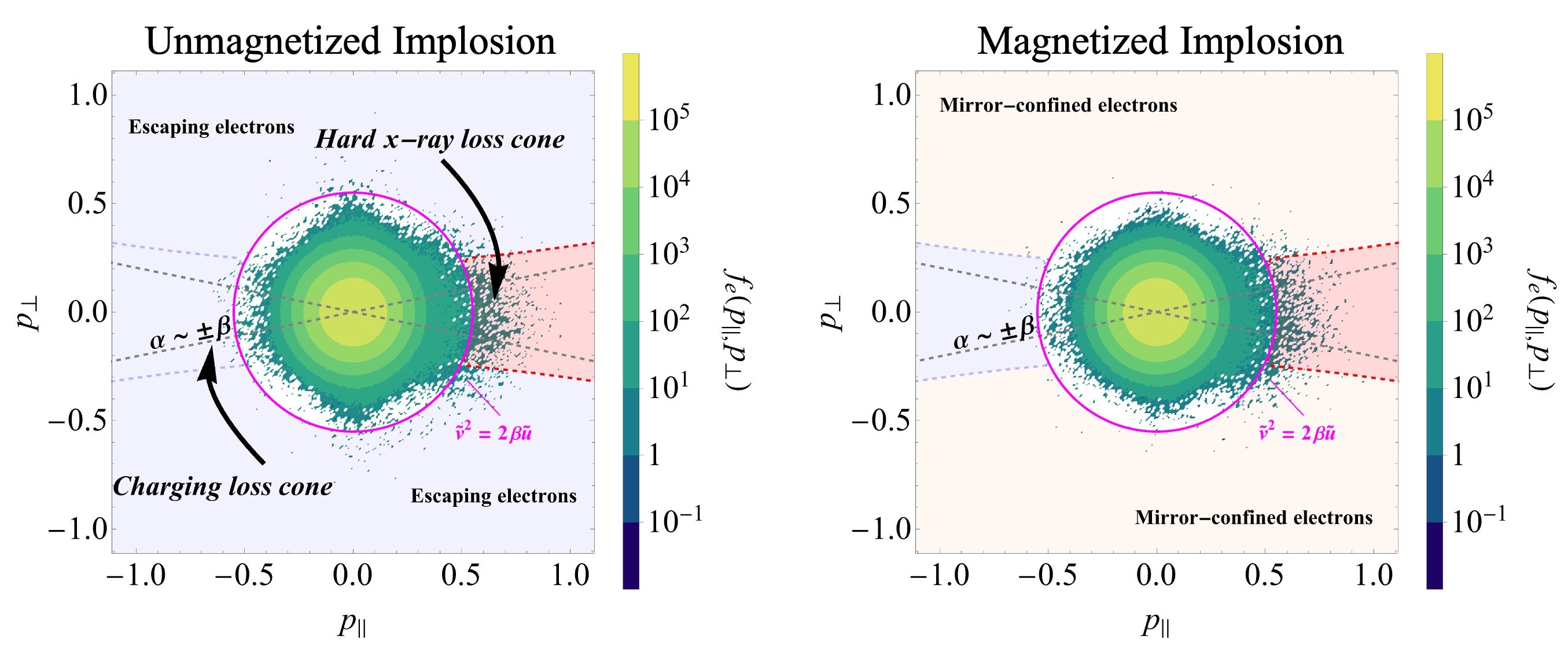}
\caption{Electron momentum distribution functions at $8.3$ \si{\pico\second} from unmagnetized (left) and magnetized (right) PIC simulations. The momentum distribution functions exhibit no significant differences between simulations with and without a magnetic field. The parallel and perpendicular directions are defined such that $p_\parallel > 0$ is directed towards the shell surface and $p_\perp$ is directed azimuthally. }
\label{fig:ElectronMomentumPhaseSpace}
\end{figure}

Fig \ref{fig:electronEnergySpectrum} shows the electron distribution for both the magnetized and unmagnetized simulations $8.3$ \si{\pico\second} after the onset of the laser. Evidently, the number of hot electrons is essentially unchanged between simulations, i.e. application of the magnetic field does not affect the total number of hot electrons. Fig \ref{fig:ElectronMomentumPhaseSpace} illustrates that the electron distribution in $p_\parallel, p_\perp$ phase space is similarly unaffected. The increased preheat is therefore attributed to the mirror-confined electrons pitch-angle scattering into the x-ray loss-cone and then colliding with the shell. Equating the ratio of the number of electrons in the mirror-confined region and x-ray loss cone in the magnetized simulation compared to the number of electrons in just the x-ray loss cone to the preheat ratio gives a simulated preheat ratio of 2.1, comparable to the measured values.

\begin{figure}[h]
    \centering
     \includegraphics[width=\linewidth]{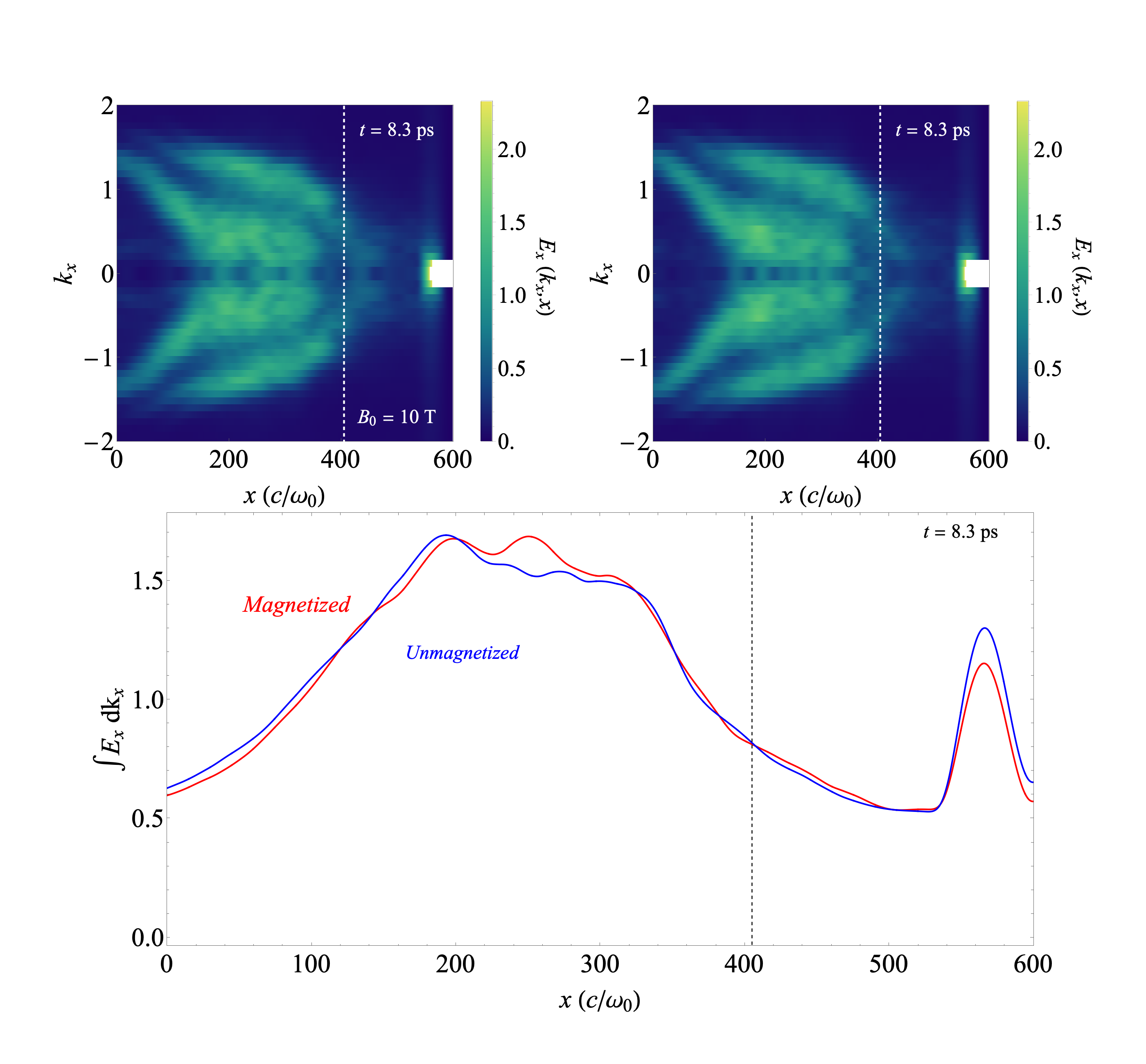}
     \caption{Top left: longitudinal electric field spectra in $k_x-x$ taken from the OSIRIS simulation with an external magnetic field present at $t = 8.3$ \si{\pico\second}. Top right) same as left but no external magnetic field is present. Bottom) Integral of top left and right taken along $k_x$ illustrating the magnetic field does not significantly affect the instability or electric field profiles. The vertical-dashed line is drawn at the location of the quarter critical density.}
     \label{fig:Exspectra_kxx}
\end{figure}

OSIRIS simulated spectra of the saturated longitudinal electric field with and without a magnetic field are shown in Fig. \ref{fig:Exspectra_kxx} at $t= 8.3$ \si{\pico\second} for comparison. The spectra were calculated by performing a short-time-Fourier transform along $x$ then averaging the results along $y$. The two spectra differ only marginally with respect to the peak amplitude at densities below critical. As such, the EPW damping and hot electron generation via the TPD instability is not substantially modified by the presence of a longitudinal magnetic field of $10 $ \si{\tesla}. Hence, modification of the TPD instability due to the external magnetic field cannot account for the observed differences in the charged-particle spectra and hard x-ray data in magnetized and unmagnetized implosions.

\bibliography{refs.bib}

\end{document}